\documentclass[conference]{IEEEtran}
\IEEEoverridecommandlockouts
\usepackage{cite}
\usepackage{amsmath,amssymb,amsfonts}
\usepackage{algorithmic}
\usepackage{graphicx}
\usepackage{array}
\usepackage{textcomp}
\usepackage{xcolor}
\usepackage{graphicx} 
\usepackage{caption} 
\usepackage{float} 
\usepackage{subcaption}
\usepackage{dblfloatfix}
\usepackage{amssymb}
\usepackage{graphicx}
\usepackage{multirow}

\def\BibTeX{{\rm B\kern-.05em{\sc i\kern-.025em b}\kern-.08em
    T\kern-.1667em\lower.7ex\hbox{E}\kern-.125emX}}
\begin{document}

\title{A Non-Zero-Sum Game Model for Optimal Cyber Defense Strategies\\
}

\author{\IEEEauthorblockN{Dongyoung Park}
\IEEEauthorblockA{\textit{Department of Computer Science} \\
\textit{Boise State University}\\
Boise, Idaho \\
youngpark@u.boisestate.edu}
\and
\IEEEauthorblockN{Gaby G. Dagher}
\IEEEauthorblockA{\textit{Department of Computer Science} \\
\textit{Boise State University}\\
Boise, Idaho \\
gabydagher@boisestate.edu}
 }

\maketitle

\begin{abstract}

In the contemporary digital landscape, cybersecurity has become a critical issue due to the increasing frequency and sophistication of cyber attacks. This study utilizes a non-zero-sum game theoretical framework to model the strategic interactions between cyber attackers and defenders, with the objective of identifying optimal strategies for both. By defining precise payoff functions that incorporate the probabilities and costs associated with various exploits, as well as the values of network nodes and the costs of deploying honeypots, we derive Nash equilibria that inform strategic decisions. The proposed model is validated through extensive simulations, demonstrating its effectiveness in enhancing network security. Our results indicate that high-probability, low-cost exploits like Phishing and Social Engineering are more likely to be used by attackers, necessitating prioritized defense mechanisms. Our findings also show that increasing the number of network nodes dilutes the attacker's efforts, thereby improving the defender's payoff. This study provides valuable insights into optimizing resource allocation for cybersecurity and highlights the scalability and practical applicability of the game-theoretic approach.
\end{abstract}


\section{Introduction}

Cybersecurity is becoming an ever-more critical concern in the digital world. The rapid advancement of technology and the concomitant rise in cyber attacks demand the development of robust and efficient security systems. Game theory provides a powerful framework for modeling and analyzing the strategic interactions between cyber attackers and defenders \cite{attiah2018game} \cite{zhang2017game}. There are various types of games, and depending on the perspective, they can be classified in several different ways \cite{gibbons1992game} \cite{owen2013game}. Particularly, non-zero-sum games allow for a more realistic representation of the cybersecurity landscape, where both attackers and defenders have their respective goals and constraints, leading to significant contributions in developing effective security strategies \cite{attiah2018game} \cite{ferdowsi2017colonel}. 

One of the key aspects of cybersecurity is understanding and mitigating the impact of exploits. An exploit refers to a piece of software, data, or sequence of commands that takes advantage of a bug or vulnerability to cause unintended behavior in software or hardware systems. Exploits are fundamental tools for attackers, enabling them to gain unauthorized access, disrupt services, or steal sensitive information. Notable examples include the EternalBlue exploit, which facilitated the infamous WannaCry ransomware attack \cite{eternalblue}, and the Log4Shell vulnerability, a zero-day exploit that allowed remote code execution in the widely-used Apache Log4j library \cite{log4shell}. Additionally, the Heartbleed vulnerability in OpenSSL highlighted the critical need for securing communication channels and protecting sensitive data \cite{heartbleed}. Other significant exploits include SQL Injection vulnerabilities such as those in Drupal \cite{sqlinjection2014} and PHP-FPM \cite{sqlinjection2019}, which can allow attackers to manipulate databases, and the Adobe Flash Player vulnerability that can be exploited through phishing attacks \cite{phishing2018}. Furthermore, the BlueKeep vulnerability in Microsoft's Remote Desktop Protocol (RDP) has been linked to ransomware attacks \cite{ransomware2019}. In a complex network environment, such as those in many enterprises, it is possible for multiple vulnerabilities to exist simultaneously, each posing significant risks if not properly mitigated. Attackers gather inside information about targeted networks using various tools and scanning techniques \cite{engebretson2013basics}, while defenders must constantly update their strategies to counter these exploits. Techniques such as cyber deception, which manipulates network interfaces to disguise the true state of the network, are crucial in this ongoing battle \cite{aslan2023comprehensive}.

This study aims to leverage non-zero-sum game theory to define the reward functions for cyber attackers and defenders, identify Nash equilibrium, and derive optimal strategies for both players. Specifically, we incorporate the probabilities and costs associated with various exploits used by attackers, the value of each network node, the costs of deploying honeypots by defenders, and the total node values. By calculating the expected payoffs at Nash equilibrium, we propose optimal strategies for each player that enhance overall cybersecurity.

In the field of cybersecurity, utilizing game theory to define payoff functions for both attackers and defenders is crucial \cite{alpcan2010network}. The primary challenge lies in creating payoff functions that accurately reflect the dynamic interactions between attackers and defenders. Attackers employ various exploits to compromise the system, while defenders allocate resources to safeguard it \cite{carroll2011game} \cite{zhang2010gpath} \cite{zhu2013game}. Our model considers several critical factors: the probabilities and costs associated with each exploit, the values assigned to network nodes based on their importance, and the costs incurred by defenders in deploying honeypots. By applying a non-zero-sum game framework, we analyze these interactions to derive Nash equilibrium. This analysis informs the optimal strategies for both attacking and defending, providing a robust approach to enhancing cybersecurity measures.

The methodology of this study involves several key steps. First, we define the reward functions for attackers and defenders, considering the assigned exploits and their costs for each node. Exploits are characterized by their usage probabilities, which represent the likelihood of an exploit being employed by an attacker, and their costs, which include both the development and deployment expenses. Next, we compute the optimal attack and defense strategies based on the reward functions. Using mathematical modeling and optimization techniques, we then identify Nash equilibrium. Finally, we conduct simulations to calculate the expected payoffs for each strategy, compare execution times with and without heuristic methods, and analyze the model's scalability.

\subsection{Contributions}
The contributions of this paper are as follows:
\begin{itemize}
    \item We introduce a novel non-zero-sum game theory model for analyzing static target identification and allocation in cybersecurity to allow for a practical representation of the interactions between attackers and defenders, facilitating the derivation of optimal strategies.
    \item We develop reward functions for defender and attacker that incorporate the probabilities and costs associated with various exploits, as well as the values of network nodes and the costs of deploying honeypots to enable a detailed characterization of Nash equilibrium.
    \item We formulate optimal mixed strategies for both attackers and defenders, demonstrating how these strategies can be adjusted based on different network configurations and node values to maximize the defender’s payoff.
    \item We provide experimental results demonstrating the robustness and effectiveness of our approach in identifying strategies that enhance defender payoffs against attacker payoffs, highlighting the scalability and practical applicability of the game-theoretic framework in real-world cybersecurity contexts.
\end{itemize}

The rest of this paper is organized as follows. In Section~\ref{sec:Related work}, we review existing literature, highlighting the differences and contributions of this study. Section~\ref{sec:Methodology} offers a detailed explanation of the non-zero-sum game model, including payoff functions, node values, and cost settings, and presents the mathematical modeling and optimization methods used to derive optimal strategies. Section~\ref{sec:Case Study} provides a practical application of our model, demonstrating its effectiveness in real-world scenarios by defining various exploits, simulating multiple iterations, and analyzing outcomes to determine optimal strategies and Nash equilibrium for both attackers and defenders. Section~\ref{sec:Experiment Evaluation} discusses the simulation results, including a comparison of execution times and scalability analysis. Finally, Section~\ref{sec:Conclusion} summarizes the findings and discusses future research directions.
\section{Related Work} \label{sec:Related work}

The application of game theory to cybersecurity has been extensively studied, providing valuable frameworks for modeling the strategic interactions between cyber attackers and defenders. Wang et al. \cite{7866199} provide a comprehensive review of how game theory has been employed to address different security application scenarios, such as cyber attack-defense analysis and security assessment. Similarly,  Do et al. \cite{do2017game} explore various attack and defense scenarios, offering a thorough overview of game theory applications for cybersecurity and privacy issues.

In addition, Tsai et al. \cite{tsai2012security} discuss the use of security games to control the spread of contagious attacks, such as worms, offering strategies to defend against such threats in networked environments. They model this problem as a zero-sum dynamic game where the attacker aims to maximize the spread of influence (e.g., a worm) while the defender aims to minimize it. Zhu et al. \cite{zhu2011dynamic} propose a dynamic secure routing game framework to combat jamming attacks in distributed cognitive radio (CR) networks. The game model is a stochastic multi-stage zero-sum game based on the directional exploration of ad hoc on-demand distance vector (AODV) algorithms. 

Game theory has been extensively applied to address the security challenges in unmanned systems, providing a robust framework for modeling and analyzing adversarial scenarios. Jahan et al. \cite{9283823} present a new architecture for modeling attacks on autonomous systems, utilizing a non-cooperative non-zero sum game to derive optimal strategies for maintaining a secure system state. Similarly, Sanjab et al. \cite{sanjab2017prospect} present a static zero-sum game model to address the cyber-physical security challenges in drone delivery systems. Their work incorporates prospect theory to model the subjective behavior of both the vendor and the attacker, capturing the subjective perception of attack success probabilities and their disparate valuations of delivery times.

Cyber deception strategies have also garnered significant attention in recent years as a proactive means of defending against sophisticated cyber threats. Basak et al. \cite{basak2021scalable} propose a Stackelberg game model and a scalable solution approach that uses both exact algorithms and heuristics to enhance the early detection and identification of stealthy attackers in cyber-physical networks. The integration of deception techniques, such as honeypot deployment, plays a critical role in forcing attackers to reveal their strategies and identities, thereby improving the overall security posture of the network. Additionally, another study \cite{inbook} presents a new approach to allocate honeypots over the attack graph using two-person zero-sum strategic form games to reduce the attacker’s capability to reach a target node. Various cyber deception games have also been utilized for different purposes, highlighting the versatility and effectiveness of these strategies in enhancing cybersecurity defenses.

\begin{table*}[h] \label{table:comparison with related work}
\centering
\caption{Comparison of our work with related works in the areas of: Game Model, Solution, Application.}
\label{table}
\begin{tabular}{|>{\centering\arraybackslash}m{3cm}|>{\centering\arraybackslash}m{3.5cm}|>{\centering\arraybackslash}m{2cm}|>{\centering\arraybackslash}m{2cm}|>{\centering\arraybackslash}m{2cm}|>{\centering\arraybackslash}m{3cm}|}
\hline

\hline
\multirow{2}{*}{\textbf{Work}} & \multirow{2}{*}{\textbf{Game Model}} & \multicolumn{3}{c|}{\textbf{Solution}} & \multirow{2}{*}{\textbf{Application}} \\ \cline{3-5}
 & & \textbf{Nash Equilibrium} & \textbf{Saddle-point Equilibrium} & \textbf{Markov Equilibrium} & \\ \hline
\multirow{1}{*}{Zhu et al. \cite{zhu2011dynamic}} & \multirow{1}{*}{Dynamic zero-sum game} &  & \checkmark & & Distributed cognitive radio (CR) networks \\ \hline
\multirow{1}{*}{Jahan et al. \cite{9283823}} & \multirow{1}{*}{Static non-zero-sum game} & \checkmark &  & & Autonomous Systems \\ \hline

\multirow{1}{*}{Basak et al. \cite{basak2021scalable}} & \multirow{1}{*}{Stackelberg game} & \checkmark &  & & Cyber Deception \\ \hline
\multirow{1}{*}{Park et al. \cite{park2014trusted}} & \multirow{1}{*}{Markovian game} &  &  & \checkmark & Online Social Network \\ \hline
\multirow{1}{*}{Zhang and Zhu. \cite{zhang2015secure}} & \multirow{1}{*}{Dynamic non-zero-sum game} & \checkmark  & \checkmark  &  & Distributed Machine Learning \\ \hline
\hline
\multirow{1}{*}{\textbf{This Paper}} & \multirow{1}{*}{Static non-zero-sum game} & \checkmark &  &  & Cyber Deception \\ \hline
\end{tabular}
\end{table*}

Furthermore, Park et al. \cite{park2014trusted} introduce a zero-sum Markov game model to address the data management challenges in Online Social Network (OSN) services. They aim to manage data optimally by sharing desired information while protecting private data, whereas the attacker attempts to expose private information or disrupt the sharing process. Another significant contribution by Zhang et al. \cite{zhang2015secure} explores a Non-zero sum game model to secure and enhance the resilience of distributed machine learning algorithms in adversarial environments. Their objective is to minimize the classification error by training a Support Vector Machine (SVM) model across a network of distributed nodes.

Table \ref{table} compares our work to existing research in terms of game models, solutions, and applications within various cybersecurity contexts. While many related works utilize different game models, they do not specifically address the application of non-zero-sum games for cyber deception. Our approach stands out by focusing on non-zero-sum games, offering a more nuanced and practical framework for cyber deception. This allows for a deeper understanding and mitigation of the impact of exploits, enhancing cybersecurity measures.

\begin{table}[h]
\centering
\caption{Summary of Notations}
\begin{tabular}{m{1cm}|m{7cm}}
\hline
\textbf{Notations} & \textbf{Description} \\ \hline \hline
$\mathcal{V}$ & The set of nodes \\ \hline
$\mathcal{E}$ & The set of edges\\ \hline
$n$ & The number of nodes \\ \hline
$e_{u,v}$ & An edge from node \(u\) to node \(v\) \\ \hline
$\phi_{u,v}$ & An exploit used on edge \(e_{u,v}\) \\ \hline
$d_i$ & A strategy of the defender \\ \hline
$a_j$ & A strategy of the attacker \\ \hline
$\mathcal{N}$ & The set of players \\ \hline
$\mathcal{S}$ & The strategy space\\ \hline
$\mathcal{R}$ & Reward function representing the payoffs for the players\\ \hline
$u$ & The entry node \\ \hline
$v$ & The target node \\ \hline
$V$ & Total value of all nodes connected with the entry node \(u\). \\ \hline
$V_v$ & The value of damage by the node \( v \) that was successfully attacked. \\ \hline
$C_H$ & The cost of deploying a honeypot on the edge. \\ \hline
$P_{\phi_{u, v}}$ & The probability of any exploit used by the attacker from node \( u \) to node \( v \). \\ \hline
$C_{\phi_{u, v}}$ & The cost of the exploit used to attack from node \( u \) to node \( v \). \\ \hline
$X$ & The set of all mixed strategies for the defender \\ \hline
$Y$ & The set of all mixed strategies for the attacker \\ \hline
$r^D_{i,j}$ & The defender's payoff when choosing strategy \(i\) against the attacker's strategy \(j\). \\ \hline
$r^A_{i,j}$ & The attacker's payoff when choosing strategy \(i\) against the defender's strategy \(j\). \\ \hline
$EU$ & The expected payoff \\ \hline

\end{tabular}
\end{table}
\section{Methodology}\label{sec:Methodology}
In this section, we demonstrate the formulation of a defender-attacker security game as a non-zero-sum game. Rational players aim to maximize their own rewards, which leads us to introduce our game model, payoff matrices, the players' optimal actions, and the calculation of the Nash Equilibrium. 

\subsection{Attack Graph}
\begin{figure}[h!] 
    \centering
    \includegraphics[width=\linewidth]{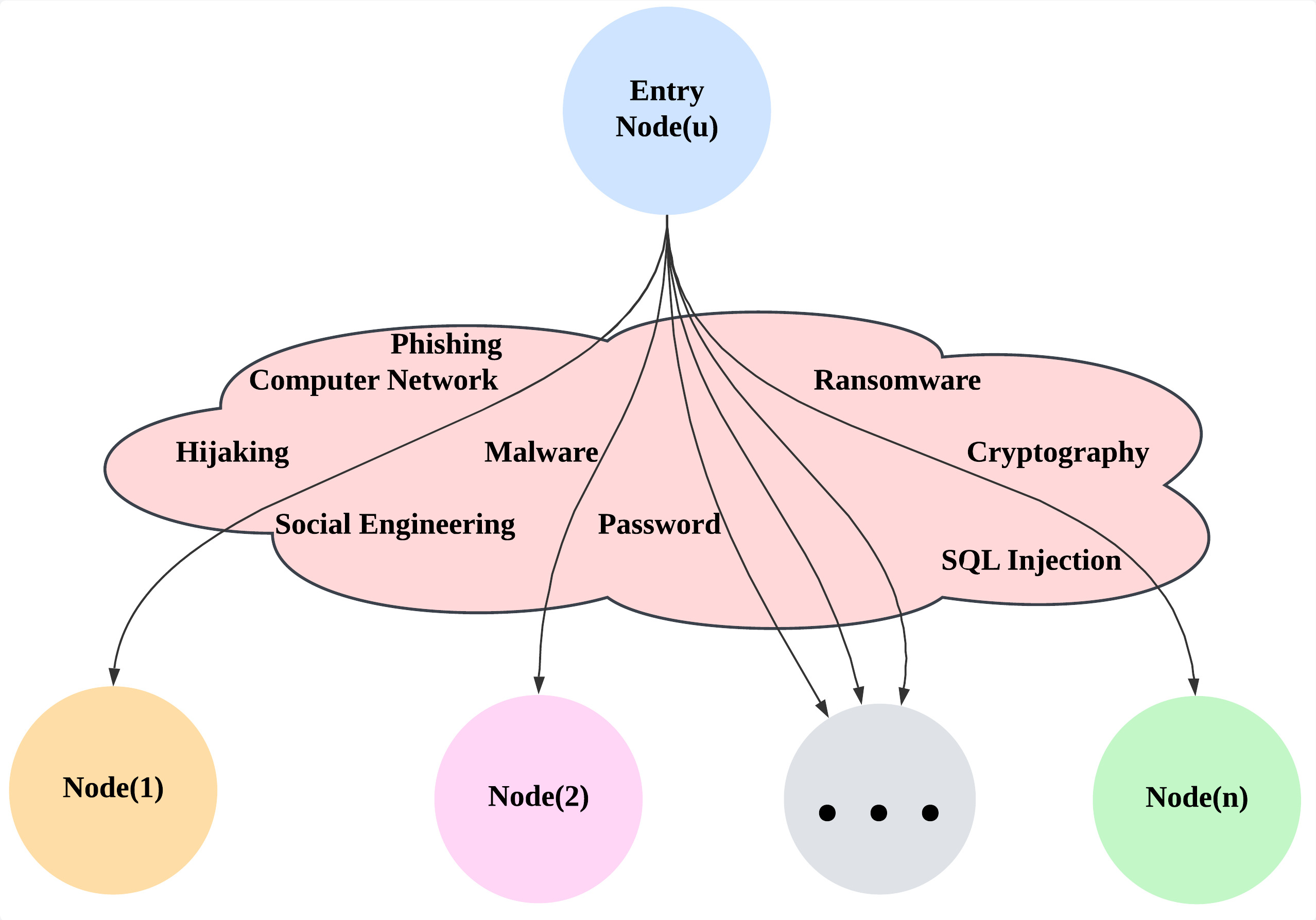} 
    \caption{Attack Graph} 
    \label{fig:attack-graph} 
\end{figure}

We define an attack graph consisting of $n$ nodes, represented as $G(\mathcal{V}, \mathcal{E})$, where $n = |\mathcal{V}|$, as shown in Figure~\ref{fig:attack-graph}. In this graph, each node $v$ has a value $V_v$, representing its importance within the network. Each edge $e_{u,v} \in \mathcal{E}$ signifies potential paths that an attacker can use to move from entry node $u$ to target node $v$ by exploiting various vulnerabilities. Each edge can have one exploit or multiple exploits, each with its own probability of success and associated cost. The attacker aims to maximize their reward by choosing the optimal strategy, which involves selecting the exploit on each edge that provides the highest reward, considering the probabilities of success and the associated costs.

In our model, we consider the attacker to have knowledge of the value associated with each node. This assumption is realistic, as attackers can often gather partial information about the network's configuration through reconnaissance activities such as network scanning.

\subsection{Game Model}
We consider a two-player non-zero-sum security game represented by \( \mathcal{G} = \langle \mathcal{N}, \mathcal{S}, \mathcal{R} \rangle \), where \( \mathcal{N} = \{D, A\} \) represents the two players: Player D is a defender and Player A is an attacker. \( \mathcal{S} = \{d_i, a_j | i, j \in \{1, 2, \ldots, n\} \} \) is the action set of all the possible strategies of the players. The defender action space is to place the honeypot on the edge or no-allocation. The attacker action is to attack node or no attack. The reward function for each player are given by \( \mathcal{R} \).

Player A attempts to exploit nodes in the network, where the set of available exploits and their probabilities are defined. Player D defends these nodes with different values and associated honeypot costs. Each player's strategy influences their respective payoffs, which are computed based on the success of the attacks and the effectiveness of the defenses.

\subsubsection{Defender's Reward Function}
The defender's reward function for defending an edge from \( u \) to \( v \) is given by:
\begin{equation}
\small
R(D, u, v) = 
\begin{cases} 
V - C_H & \text{if } A, D \rightarrow v \\
V - V_v \cdot P_{\phi_{u,v}} - C_H & \text{if } A \rightarrow v, D \not\rightarrow v 
\end{cases}
\normalsize
\label{eq:defender_reward}
\end{equation}

The defender's reward function \eqref{eq:defender_reward} reflects the net benefit of defending against attacks. If both the attacker and defender target the same node \( v \), the reward is calculated as the total value of all nodes connected to the entry node \( u \), minus the cost of deploying a honeypot on the edge. If the defender does not target node \( v \) but the attacker does, the reward is calculated as the total value \( V \) minus the product of the probability of a successful exploit and the value of node \( v \), minus the cost of deploying the honeypot.

\subsubsection{Attacker's Reward Function}
The attacker's reward function for targeting node \( v \) from node \( u \), \( R(A, u, v) \), is defined as:
\begin{equation}
\small
R(A, u, v) = 
\begin{cases} 
- C_{\phi_{u,v}} & \text{if } A, D \rightarrow v \\
V_v \cdot P_{\phi_{u,v}} - C_{\phi_{u,v}} & \text{if } A \rightarrow v, D \not\rightarrow v 
\end{cases}
\normalsize
\label{eq:attack_reward}
\end{equation}

The attacker's reward function \eqref{eq:attack_reward} captures the outcome of an attack on node \( v \) from node \( u \). If both the attacker and defender target the same node \( v \), the attacker incurs the cost of the exploit without any gain. If the defender does not target node \( v \) but the attacker does, the reward for the attacker is calculated as the product of the probability of a successful exploit and the value of node \( v \), minus the cost of the exploit used.

\subsection{Payoff Matrices}
In a strategic interaction between a defender (D) and an attacker (A), each player has a set of strategies. The outcomes of these strategies are quantified in their respective payoff matrices, $\mathcal{R}^{D}$ for the defender and $\mathcal{R}^{A}$ for the attacker, based on the previously defined reward functions.

\begin{itemize}
    \item Defender's Payoff Matrix ($\mathcal{R}^{D}$): Each element $r^D_{i,j}$ represents the payoff for defending node $v_i$ against an attack on node $v_j$, calculated as:
    \begin{equation}
    r^D_{i,j} = 
    \begin{cases} 
    V - C_H & \text{if } i = j \\
    V - V_{v_j} \cdot P_{\phi_{u,v_j}} - C_H & \text{if } i \neq j 
    \end{cases}
    \end{equation}

    \item Attacker's Payoff Matrix ($\mathcal{R}^{A}$): Each element $r^A_{i,j}$ represents the payoff for attacking node $v_j$ from node $u$, calculated as:
    \begin{equation}
    r^A_{i,j} = 
    \begin{cases} 
    - C_{\phi_{u,v_j}} & \text{if } i = j \\
    V_{v_j} \cdot P_{\phi_{u,v_j}} - C_{\phi_{u,v_j}} & \text{if } i \neq j 
    \end{cases}
    \end{equation}
\end{itemize}

A bimatrix game format is utilized to represent the payoffs for both players simultaneously, where the defender's strategies are denoted by the rows and the attacker's strategies by the columns. Each cell $(i,j)$ in this bimatrix contains a payoff pair $(r^D_{i,j}, r^A_{i,j})$, representing the defender's and the attacker's payoffs for their selected strategies.

The complete bimatrix for a non-zero sum game is given by:
\[
\begin{array}{c|*{5}{c}}
    & \mathcal{S}^A_{v_1} & \mathcal{S}^A_{v_2} & \cdots & \mathcal{S}^A_{v_n} \\
    \hline
    \mathcal{S}^D_{v_1} & (r^D_{1,1}, r^A_{1,1}) & (r^D_{1,2}, r^A_{1,2}) & \cdots & (r^D_{1,n}, r^A_{1,n}) \\
    \mathcal{S}^D_{v_2} & (r^D_{2,1}, r^A_{2,1}) & (r^D_{2,2}, r^A_{2,2}) & \cdots & (r^D_{2,n}, r^A_{2,n}) \\
    \vdots & \vdots & \vdots & \ddots & \vdots \\
    \mathcal{S}^D_{v_n} & (r^D_{n,1}, r^A_{n,1}) & (r^D_{n,2}, r^A_{n,2}) & \cdots & (r^D_{n,n}, r^A_{n,n})  \\
\end{array}
\]

\subsection{Optimal Strategy}
In determining the optimal strategy for the attacker, it is crucial to consider the reward from each potential exploit. Given multiple exploits on each edge, the attacker evaluates all available exploits and selects the one that offers the highest reward. This involves calculating the reward for each exploit based on its probability of success and associated cost, and then choosing the exploit that maximizes this value. Formally, for each edge \(e_{u,v_j}\), the attacker selects the exploit \(\phi_{u,v_j}^{*}\) that satisfies:

\begin{equation}
\phi_{u,v_j}^{*} = \max_{\phi_{u,v_j}} \left( V_{v_j} \cdot P_{\phi_{u,v_j}} - C_{\phi_{u,v_j}} \right),
\end{equation}

where \( j \in \{1, 2, \ldots, n\} \).

The attacker evaluates the reward for each exploit on all edges from the starting node \(u\) to each potential target node \(v_j\) and selects the one with the highest reward. This ensures that the attacker maximizes their overall reward from attacking the network by choosing the most effective exploits for each target node.

The attacker proceeds with an attack if the reward \(R(A, u, v)\) is greater than or equal to zero. Formally, the condition for an attack to be executed is:
\begin{equation}
V_{v_j} \cdot P_{\phi_{u,v_j}} - C_{\phi_{u,v_j}} \geq 0.
\label{eq:threshold}
\end{equation}

This condition \eqref{eq:threshold} ensures that the attacker only utilizes exploits that provide a non-negative reward. By adhering to this rule, the attacker can optimize their strategy to maximize their overall reward from attacking the network.

The rationale behind this condition is grounded in the fundamental principles of rational decision-making and utility maximization. If an exploit's reward is negative, the attacker incurs a loss, making the attack counterproductive. By focusing only on exploits with non-negative rewards, the attacker ensures that each action taken contributes positively to their overall objective.

To define the optimal strategy, the attacker evaluates all possible exploits on each edge and selects those that satisfy equation \eqref{eq:threshold}. The attacker then prioritizes these exploits based on their rewards, choosing the ones with the highest positive values. This approach guarantees that the attacker maximizes their total reward while minimizing unnecessary costs.

For the defender, understanding this optimal strategy is crucial for effective defense planning. By anticipating the attacker's behavior, the defender can allocate resources more efficiently, placing honeypots on nodes that are most likely to be targeted.

\subsection{Determining Nash Equilibrium}

The Nash Equilibrium (NE) is a set of strategies where no player can benefit by unilaterally changing their strategy. In our context, the mixed strategies for the defender and attacker are represented by probability distributions \(X\) and \(Y\), respectively.

For the defender's expected payoff, let \(X\) be a set of all mixed strategies with the probability \(x_i\) of the defender's action, such that:
\begin{equation}
X = [x_1, x_2, \ldots, x_n],
\end{equation}
where \(n\) represents the number of defender's actions.

The defender's strategy must satisfy the following constraints:
\begin{equation}
\sum_{i=1}^{n} x_i = 1, \quad x_i \geq 0 \quad \forall i \in \{1, 2, \ldots, n\}
\label{eq:defender_constraint}
\end{equation}

Constraint \eqref{eq:defender_constraint} ensures that x is a probability distribution. Similarly, for the attacker's expected payoff, let \(Y\) be a set of all mixed strategies with the probability \(y_j\) of the attacker's action, such that:
\begin{equation}
Y = [y_1, y_2, \ldots, y_n],
\end{equation}
where \(n\) represents the number of attacker's actions.

The attacker's strategy must satisfy the following constraints:
\begin{equation}
\sum_{j=1}^{n} y_j = 1, \quad y_j \geq 0 \quad \forall j \in \{1, 2, \ldots, n\}
\label{eq:attacker_constraint}
\end{equation}
Constraint \eqref{eq:attacker_constraint} ensures that y is a probability distribution.

The expected payoff for the defender when both players use mixed strategies is given by:
\begin{equation}
EU_d = X^T \mathcal{R}^{D} Y,
\end{equation}
where $\mathcal{R}^{D}$ is the defender's payoff matrix.

Similarly, the expected payoff for the attacker is given by:
\begin{equation}
EU_a = X^T \mathcal{R}^{A} Y,
\end{equation}
where $\mathcal{R}^{A}$ is the attacker's payoff matrix.

The algorithm involves the following steps:

\subsubsection{Define the Strategy}
Construct the strategy for both the defender and attacker. These strategies represent all possible mixed strategies that satisfy the probability constraints mentioned above.

\subsubsection{Compute the Payoff }
For each combination of mixed strategies, calculate the expected payoffs for both players. This involves evaluating the expected utility functions $EU_d$ and $EU_a$ for all strategy pairs $(X, Y)$.

\subsubsection{Linear Program}
The Linear Program involves solving the following optimization problems:
\begin{equation}
\max_{X} EU_d \quad \text{subject to } \sum_{i=1}^{n} x_i = 1, \quad x_i \geq 0
\end{equation}
\begin{equation}
\max_{Y} EU_a \quad \text{subject to } \sum_{j=1}^{n} y_j = 1, \quad y_j \geq 0
\end{equation}

These optimization problems can be solved using various algorithms such as the Lemke-Howson Algorithm \cite{lemke1964equilibrium}, Simpson's Algorithm \cite{myerson1978refinements}, Support Enumeration Algorithm \cite{dickhaut1993program}, Iterative Methods \cite{brown1951iterative}, and Polytope Algorithm \cite{blumrosen2007algorithmic}. In this research, we employ the polytope algorithm, which is particularly useful for dealing with high-dimensional polytopes representing the set of feasible mixed strategies. This method allows us to identify points within the polytope that satisfy the equilibrium conditions, thus effectively finding the Nash Equilibrium.

\subsubsection{Identify Nash Equilibrium}
The Nash Equilibrium corresponds to the set of mixed strategies $(X^*, Y^*)$ where both players are playing their best responses. This is identified by finding fully labeled vertices, which indicate that no player can improve their expected payoff by unilaterally changing their strategy.

To solve for the optimal mixed strategies \(X^*\) and \(Y^*\), we set up and solve the following system of linear equations:
\begin{equation}
EU_{d, v_1} = EU_{d, v_2} = \ldots = EU_{d, v_{n}}
\end{equation}
\begin{equation}
EU_{a, v_1} = EU_{a, v_2} = \ldots = EU_{a, v_{n}}
\end{equation}

These equations ensure that the expected payoff is the same for all pure strategies within the mixed strategy. The solutions are subject to the probability constraints.

Once the optimal mixed strategies \(X^*\) and \(Y^*\) are found, the optimal expected payoffs for the defender and the attacker can be calculated as follows:

For the defender:
\begin{equation}
EU_d^* = X^{*T} \mathcal{R}^{D} Y^*
\end{equation}

For the attacker:
\begin{equation}
EU_a^* = X^{*T} \mathcal{R}^{A} Y^*
\end{equation}

The resulting mixed strategies $(X^*, Y^*)$ form a Nash Equilibrium for the defender-attacker security game. This equilibrium ensures that neither player can unilaterally improve their payoff by changing their strategy, thus providing a stable and optimal solution for both players in the context of the security game.

\section{Case Study}\label{sec:Case Study}
In this section, we present the case study of our defender-attacker security game. We define various exploits with specific costs and success probabilities, simulate multiple iterations, and analyze the outcomes to determine the optimal strategies and Nash equilibrium for both players.

\subsection{Setup}
In this study, we simulate a defender-attacker security game by defining a set of exploits with corresponding costs and success probabilities, reflecting several security scenarios. Table \ref{table:exploits} details these exploits.

\begin{table}[h]
\centering
\caption{Exploits, Costs, and Success Probabilities}
\label{table:exploits}
\begin{tabular}{c c c c}
\hline
\textbf{Type} & \textbf{Exploit} & \textbf{Success Probability} & \textbf{Cost} \\ \hline \hline
Phishing & $\phi_{1}$ & 0.7 & 3 \\ \hline
Malware & $\phi_{2}$ & 0.4 & 5 \\ \hline
Ransomware & $\phi_{3}$ & 0.3 & 8 \\ \hline
Social Engineering & $\phi_{4}$ & 0.8 & 4 \\ \hline
SQL Injection & $\phi_{5}$ & 0.6 & 7 \\ \hline
Cryptography & $\phi_{6}$ & 0.2 & 10 \\ \hline
\end{tabular}
\end{table}

The honeypot cost is set at 7. We consider three different scenarios with varying numbers of nodes: 2, 3, and 4 nodes. The configurations and values of these nodes are as follows:
\begin{itemize}
    \item \textbf{2 Nodes:} \{(A, 30), (B, 20)\}
    \item \textbf{3 Nodes:} \{(A, 25), (B, 15), (C, 10)\}
    \item \textbf{4 Nodes:} \{(A, 20), (B, 15), (C, 10), (D, 5)\}
\end{itemize}

For each node configuration, the simulation proceeded through the following steps:
\begin{enumerate}
\item \textbf{Random Assignment of Exploits:} Exploits are randomly assigned to nodes, with each node receiving between 1 and 4 exploits.
\item \textbf{Calculation of Strategies:} Optimal attack and defense strategies for each node are determined based on the assigned exploits, node values, and associated costs.
\item \textbf{Construction of Payoff Matrices:} Payoff matrices for both the defender and the attacker are constructed using the calculated strategies.
\item \textbf{Nash Equilibrium Analysis:} Nash equilibria are analyzed. For each equilibrium, expected payoffs are calculated. This process is repeated for 1000 iterations, and the average payoffs are computed to assess overall performance.
\end{enumerate}

\subsection{Optimal Mixed Strategies}
Table \ref{table:mixed_strategies} shows the average Nash Equilibrium strategies over the 1000 simulations, with fractions rounded to two digits.

\begin{table}[h]
\centering
\caption{Optimal Mixed Strategies with Heuristics}
\label{table:mixed_strategies}
\begin{tabular}{c c c}
\hline
\textbf{Node Number} & \textbf{Defender (X*)} & \textbf{Attacker (Y*)} \\ \hline \hline
2 & [0.57, 0.43] & [0.42, 0.58] \\ \hline
3 & [0.55, 0.31, 0.14] & [0.26, 0.35, 0.39] \\ \hline
4 & [0.50, 0.34, 0.15, 0.01] & [0.26, 0.31, 0.36, 0.07] \\ \hline
\end{tabular}
\end{table}

\begin{table}[h]
\centering
\caption{Optimal Mixed Strategies without Heuristics}
\label{table:mixed_strategies_no_heuristic}
\begin{tabular}{c c c}
\hline
\textbf{Node Number} & \textbf{Defender (X*)} & \textbf{Attacker (Y*)} \\ \hline \hline
2 & [0.58, 0.42] & [0.41, 0.59] \\ \hline
3 & [0.56, 0.31, 0.13] & [0.26, 0.36, 0.38] \\ \hline
4 & [0.48, 0.34, 0.16, 0.02] & [0.26, 0.32, 0.33, 0.09] \\ \hline
\end{tabular}
\end{table}

The defender's mixed strategies show a clear trend of prioritizing nodes based on their values. Higher value nodes receive a larger share of the defender's resources, reflecting the need to protect the most critical assets in the network. As the number of nodes increases, the defender's resources are more spread out, but the allocation still reflects the relative importance of each node. The attacker's strategies are influenced by the perceived vulnerabilities and rewards associated with each node. In scenarios with more nodes, the attacker tends to focus on nodes that are less defended or offer higher payoffs. This indicates an adaptive strategy where the attacker aims to exploit the weakest links in the network.

Comparing the optimal mixed strategies with and without heuristics, it is evident that the differences are minimal. These subtle differences suggest that the use of heuristics does not significantly impact the determination of optimal mixed strategies. Both heuristic and non-heuristic approaches yield similar allocations of resources and focus points for both defenders and attackers.

\subsection{Optimal Expected Payoffs}
Table \ref{table:expected_payoffs} presents the optimal expected payoffs of the defender and the attacker for different node configurations, with fractions rounded to two digits.

\begin{table}[h]
\centering
\caption{Optimal Expected Payoffs with Heuristics}
\label{table:expected_payoffs}
\begin{tabular}{c c c}
\hline
\textbf{Node Number} & \textbf{Defender's Expected} & \textbf{Attacker's Expected} \\ & \textbf{Payoff (EU\textsubscript{d}*)} & \textbf{Payoff (EU\textsubscript{a}*)} \\ \hline \hline
2 & 39.02 & 3.10 \\ \hline
3 & 39.95 & 2.30 \\ \hline
4 & 40.24 & 1.93 \\ \hline
\end{tabular}
\end{table}

\begin{table}[h]
\centering
\caption{Optimal Expected Payoffs without Heuristics}
\label{table:expected_payoffs_no_heuristic}
\begin{tabular}{c c c}
\hline
\textbf{Node Number} & \textbf{Defender's Expected} & \textbf{Attacker's Expected} \\ & \textbf{Payoff (EU\textsubscript{d}*)} & \textbf{Payoff (EU\textsubscript{a}*)} \\ \hline \hline
2 & 39.47 & 2.64 \\ \hline
3 & 40.19 & 2.13 \\ \hline
4 & 40.49 & 1.78 \\ \hline
\end{tabular}
\end{table}

The results indicate that an increase in the number of nodes leads to a higher optimal expected payoff for the defender (EU\textsubscript{d}*), whereas the attacker’s optimal expected payoff (EU\textsubscript{a}*) decreases. This trend suggests that increasing the number of nodes in the network provides a strategic advantage to the defender. More nodes dilute the attacker's efforts, making it harder to achieve successful attacks and resulting in higher costs for each successful exploit. These trends are observed in both heuristic and non-heuristic scenarios, but the heuristic approach shows a slight edge in efficiency and effectiveness.

\section{Experiment Evaluation}\label{sec:Experiment Evaluation}
In this section, we evaluate the effectiveness and robustness of our proposed defender-attacker security game model through a series of experiments. We conduct a detailed analysis of the optimal strategies and expected payoffs for both defenders and attackers under various scenarios. The experiments include examining the frequency of exploit usage, the impact of exploit success probabilities, the influence of honeypot costs, and the effect of varying exploit costs on the expected payoffs. Additionally, we assess the scalability of our model by comparing the computational run-times between heuristic and non-heuristic approaches.

\subsection{Exploit Usage Frequency Across Scenarios}
We analyze the frequency of each exploit being used as the optimal strategy across different scenarios at 4 nodes. This analysis helps in understanding the strategic choices made by the attacker in the simulation and provides insights into the most frequently utilized exploits for each node.
\begin{figure}[h]
\centering
\begin{subfigure}{0.49\linewidth}
    \includegraphics[width=\linewidth]{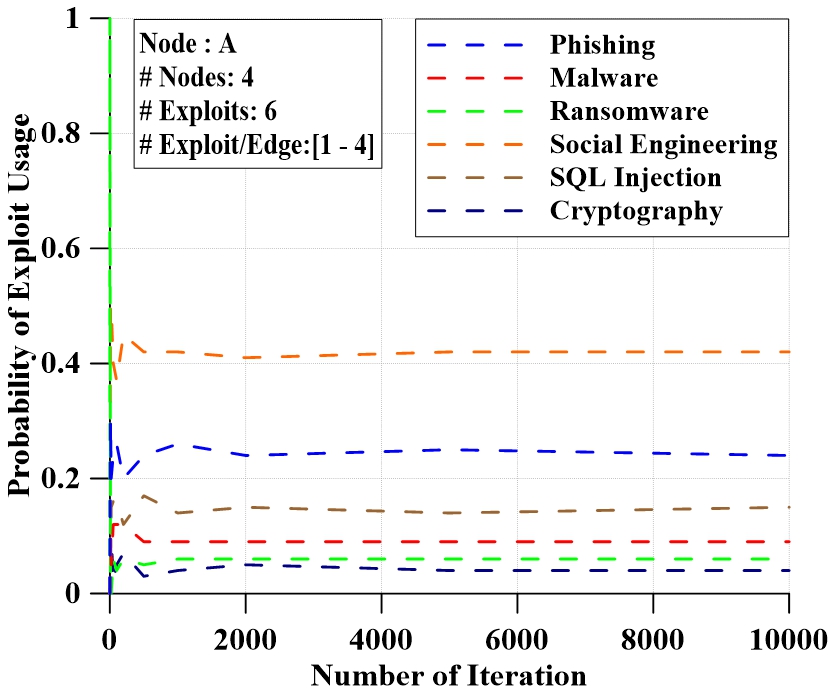}
    \caption{Node A}
\end{subfigure}
\begin{subfigure}{0.49\linewidth}
    \includegraphics[width=\linewidth]{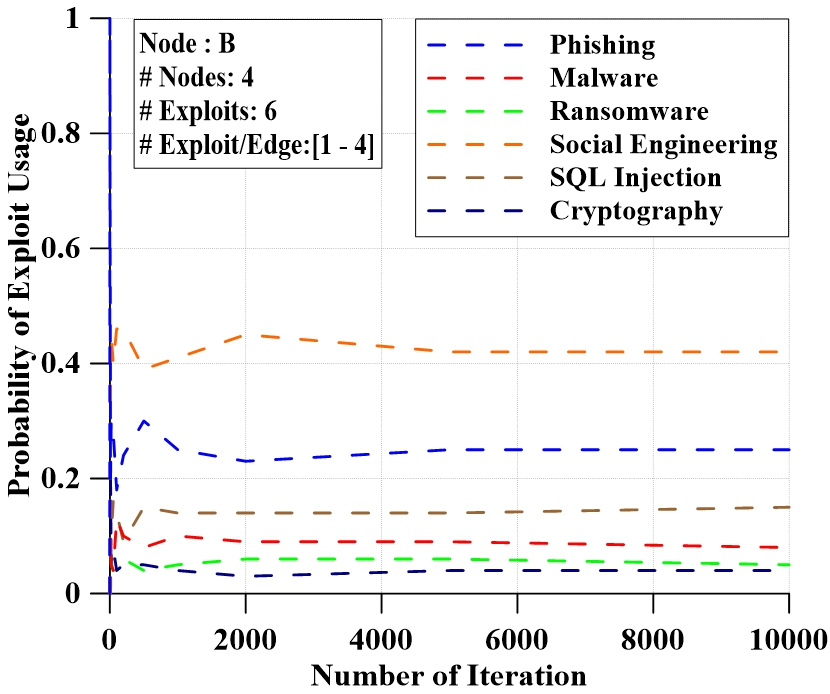}
    \caption{Node B}
\end{subfigure}
\begin{subfigure}{0.49\linewidth}
    \includegraphics[width=\linewidth]{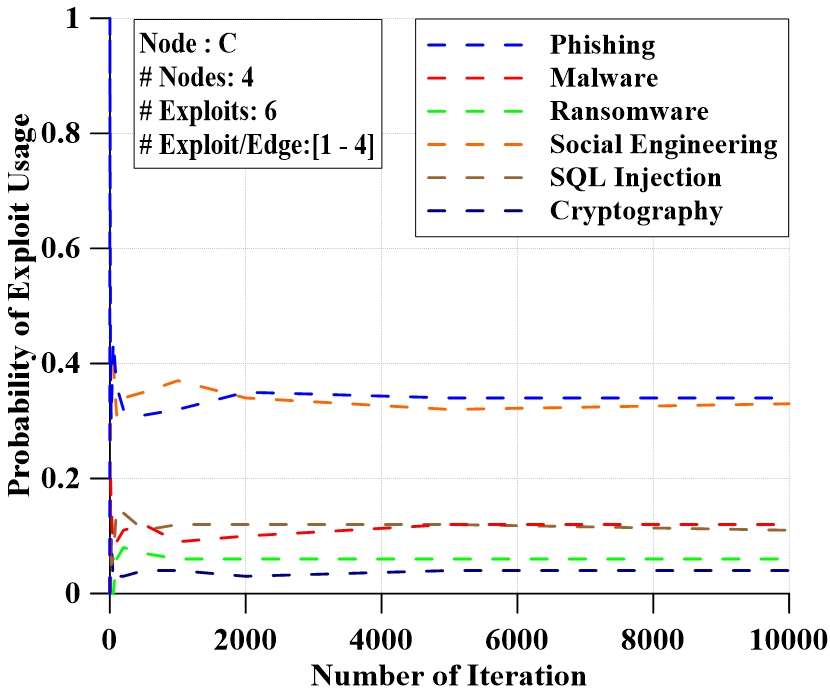}
    \caption{Node C}
\end{subfigure}
\begin{subfigure}{0.49\linewidth}
    \includegraphics[width=\linewidth]{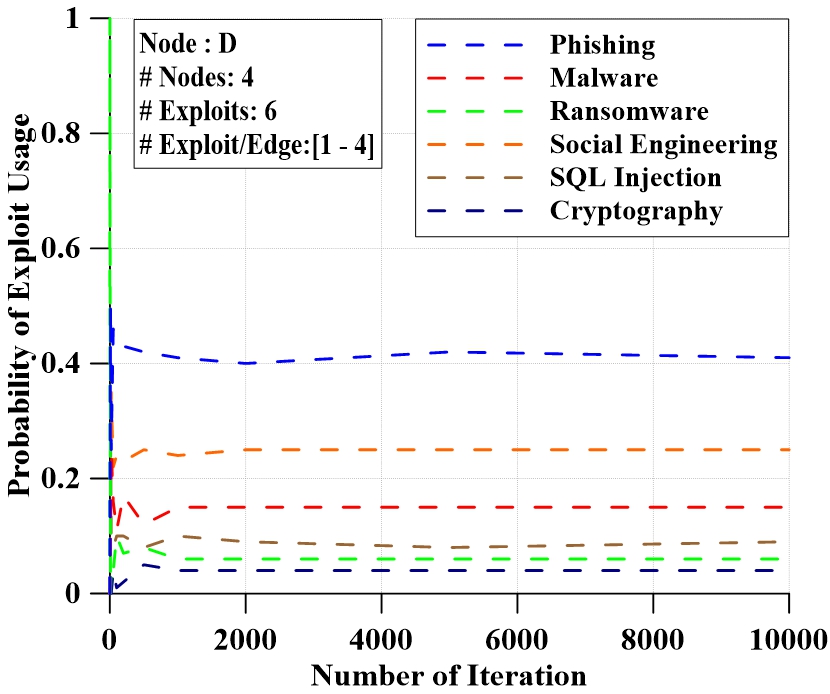}
    \caption{Node D}
\end{subfigure}
\caption{Exploit Usage Count per Node}
\label{fig:exploit usage}
\end{figure}

Figure~\ref{fig:exploit usage} illustrate the total number of times each exploit is used as the optimal strategy for nodes A, B, C, and D across all iterations. The mixed strategies used in this analysis are calculated without heuristics, ensuring a comprehensive exploration of all the strategy spaces.

Phishing and Social Engineering are used with high frequency across most nodes, indicating that these exploits offer attackers high success rates and relatively low costs. Malware and SQL Injection are also used frequently, though to a lesser extent compared to Phishing and Social Engineering. In contrast, Ransomware and Cryptography are seldom used, likely due to their higher costs or lower success probabilities.

By analyzing the probability of exploit usage based on the node values for four nodes, several trends emerge:

As the value of a node decreases, there is a noticeable shift in the types of exploits used. Higher value nodes (Nodes A and B) see a significant focus on Social Engineering and Phishing, with Social Engineering being particularly dominant. This indicates that attackers prioritize high-value nodes with exploits that maximize success rates and minimize resource usage.

In contrast, lower value nodes (Nodes C and D) exhibit a higher probability of Phishing usage, surpassing Social Engineering. Additionally, the probability of Malware usage increases in these lower value nodes compared to higher value nodes. This shift suggests that attackers opt for more cost-effective and diverse exploit strategies when targeting less critical assets.

These observations confirm that the value of a node significantly influences the probability of different exploits being used. High-value nodes attract more focused and frequent exploit attempts through Social Engineering and Phishing, whereas lower-value nodes see a shift towards more diverse and cost-effective exploits. Understanding these patterns provides crucial insights into which exploits are most attractive to attackers, allowing defenders to prioritize their defensive measures against the most frequently used exploits, thereby enhancing the overall security of the network.

\subsection{Impact of Exploit Success Probability}
The average expected payoffs for the defender and the attacker are calculated over 1000 iterations for each exploit configuration. The probability of each exploit is varied from 0 to 1 in steps of 0.1 while keeping other exploit values constant. The results are analyzed to observe the impact of varying each exploit on the expected payoffs for both the defender and the attacker.

\begin{figure*}[hbt!]
\centering
\begin{subfigure}{0.3\linewidth}
    \includegraphics[width=\linewidth,height=1\linewidth]{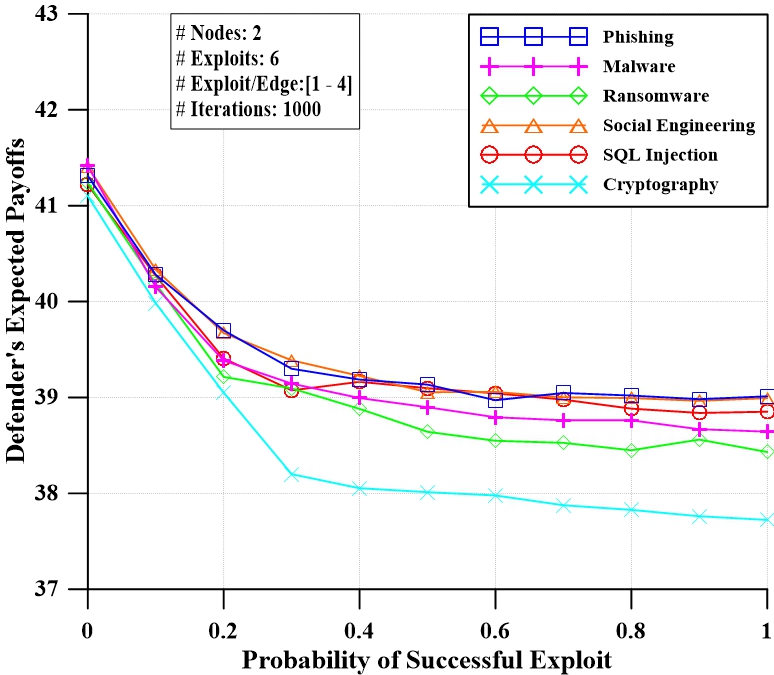}
    \caption{2 Nodes}
\end{subfigure}
\begin{subfigure}{0.3\linewidth}
    \includegraphics[width=\linewidth,height=1\linewidth]{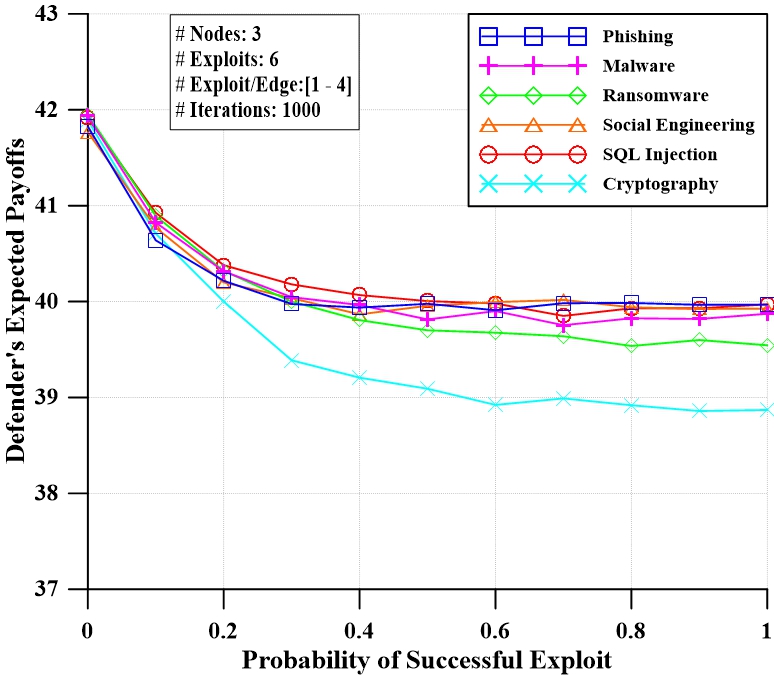}
    \caption{3 Nodes}
\end{subfigure}
\begin{subfigure}{0.3\linewidth}
    \includegraphics[width=\linewidth,height=1\linewidth]{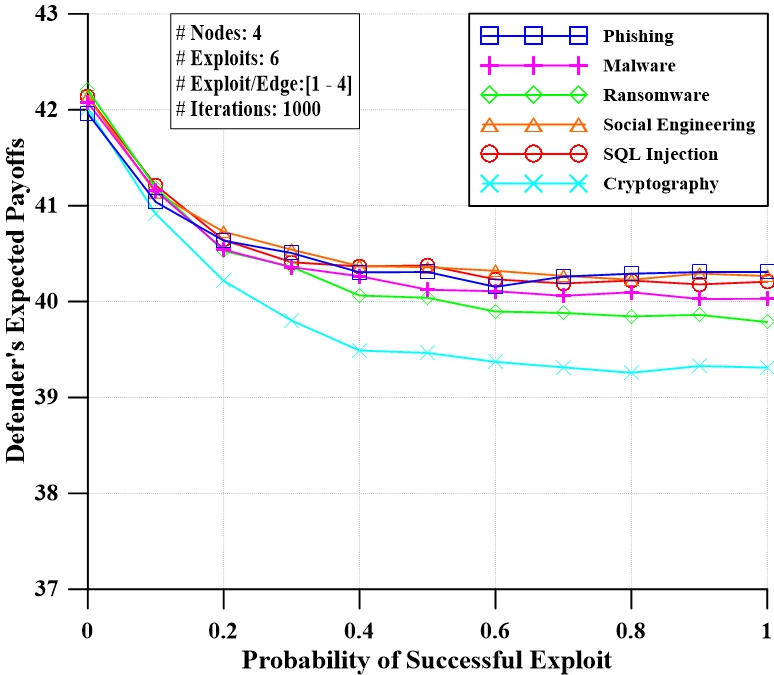}
    \caption{4 Nodes}
\end{subfigure}
\caption{Defender's Expected Payoffs with Varying Exploit Success Probabilities}
\label{fig:defender}
\end{figure*}

\begin{figure*}[hbt!]
\centering
\begin{subfigure}{0.3\linewidth}
    \includegraphics[width=\linewidth,height=1\linewidth]{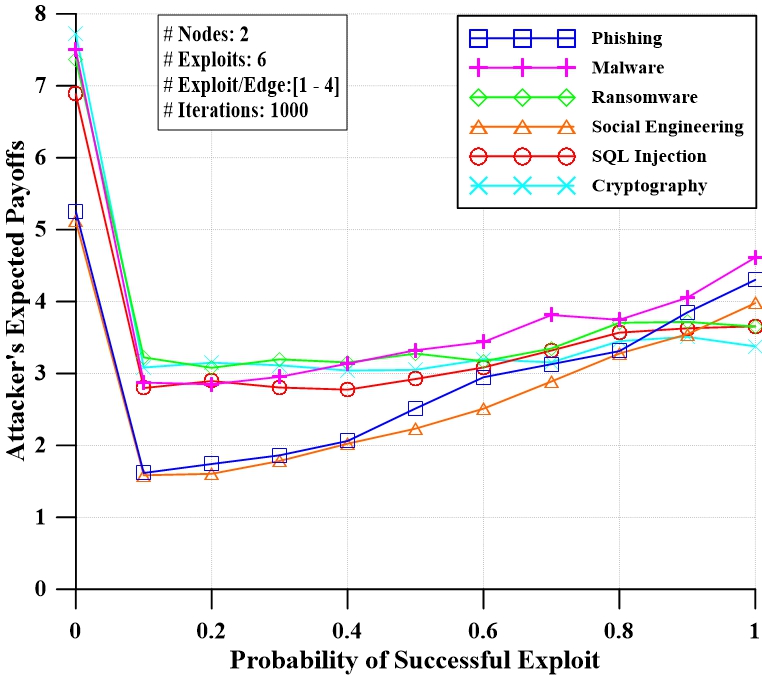}
    \caption{2 Nodes}
\end{subfigure}
\begin{subfigure}{0.3\linewidth}
    \includegraphics[width=\linewidth,height=1\linewidth]{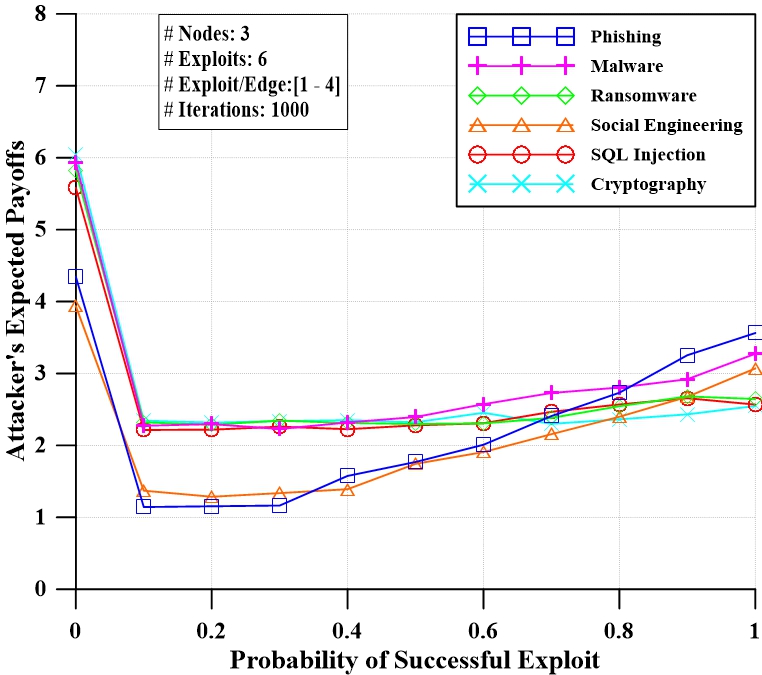}
    \caption{3 Nodes}
\end{subfigure}
\begin{subfigure}{0.3\linewidth}
    \includegraphics[width=\linewidth,height=1\linewidth]{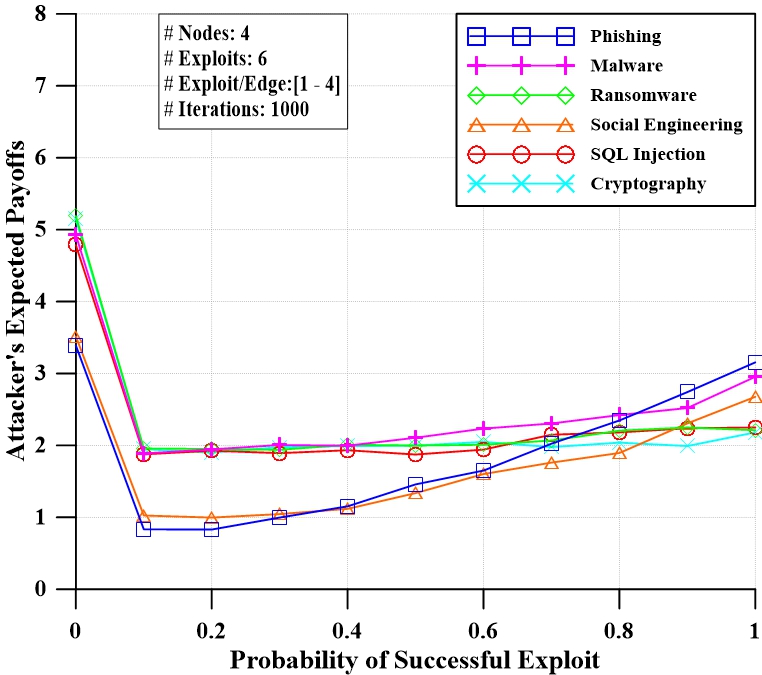}
    \caption{4 Nodes}
\end{subfigure}
\caption{Attacker's Expected Payoffs with Varying Exploit Success Probabilities}
\label{fig:attacker}
\end{figure*}

Figures~\ref{fig:defender} and ~\ref{fig:attacker} illustrate the expected payoffs for the defender and attacker as the probability of success for each exploit varies. In these simulations, the success probability of one exploit is varied while keeping the probabilities of the other five exploits constant.

Figure~\ref{fig:defender} shows that the defender's expected payoff is inversely related to the success probability of the exploits. As the success probability of high-reward exploits like Phishing and Social Engineering increases, the defender's expected payoff decreases. This trend highlights the importance of effectively defending against these high-probability, high-reward exploits to maintain a higher overall payoff.

From Figure \ref{fig:attacker}, we observe that the attacker's expected payoff decreases significantly when the probability of success for Phishing or Social Engineering is low. This is because these exploits have higher success probabilities and relatively lower costs, making them more attractive options for the attacker. As their success probabilities increase, the attacker's expected payoff also increases.

Additionally, as the number of nodes increases from 2 to 4, the expected payoffs for both the defender and attacker change accordingly. With more nodes, the defender's payoffs tend to decrease at a slower rate as the success probability of exploits increases, indicating a more resilient defense strategy in larger networks. Conversely, the attacker's payoffs show a more pronounced increase in larger networks, reflecting the higher potential rewards from successfully exploiting multiple nodes.

As the number of nodes increases, the overall trend shows that the defender's expected payoffs decrease less rapidly with increasing exploit success probabilities. This suggests that larger networks provide more opportunities for the defender to distribute resources effectively, thereby mitigating the impact of high-probability exploits. For attackers, larger networks offer more targets, and the increased expected payoffs reflect the greater likelihood of finding and exploiting vulnerabilities in some of the nodes.

The comparison between the defender and attacker shows a clear inverse relationship in their expected payoffs concerning exploit success probabilities. For defenders, maintaining high payoffs requires robust strategies to counter high-probability exploits, while attackers benefit from these high-probability exploits due to their higher success rates and lower associated costs. This interaction highlights the necessity for defenders to prioritize defenses against the most likely and potentially most damaging exploits to minimize losses effectively.

\subsection{Impact of Honeypot Costs}
\begin{figure}[H] 
    \centering
    \includegraphics[width=0.8\linewidth,height=0.8\linewidth]{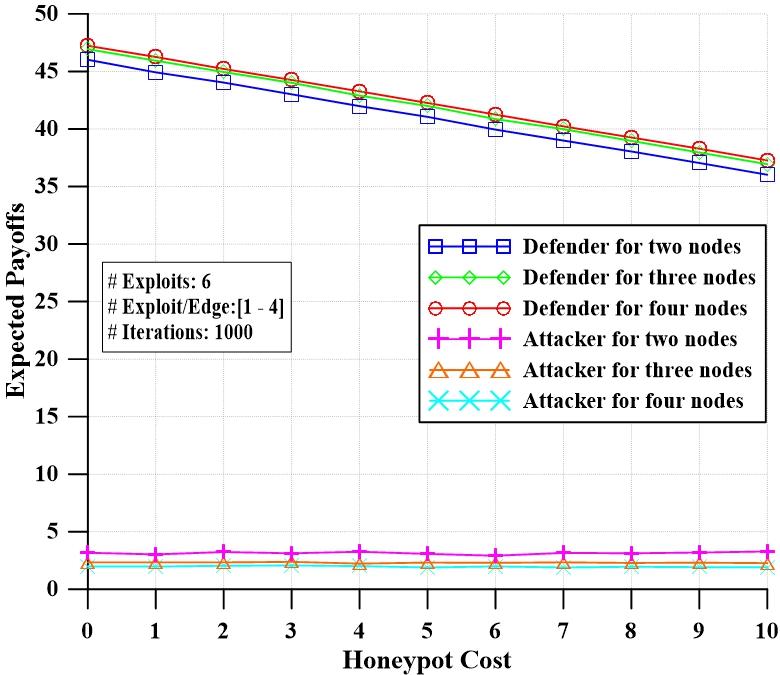} 
    \caption{Expected Payoffs with Varying Honeypot Costs} 
    \label{fig:honeypot} 
\end{figure}
Figure~\ref{fig:honeypot} provides a visual representation of the impact of different honeypot costs on the expected payoffs and 
Figure~\ref{fig:honeypot} shows how the defender's expected payoff decreases as the honeypot cost increases. 

This trend is consistent across different node configurations. Higher honeypot costs lead to a reduction in the defender's payoff because the resources that could have been used for defense are diverted to maintain the honeypots. 

This demonstrates a critical balance the defender must achieve between investing in honeypots and direct defense mechanisms.

Additionally, the attacker's expected payoff remains relatively stable regardless of the honeypot cost. This indicates that the attacker's strategy is not significantly impacted by the increased investment in honeypots, primarily because the overall defense mechanism becomes less effective with higher honeypot costs, allowing the attacker to find vulnerabilities more consistently.

\subsection{Impact of Exploit Costs}
\begin{figure}[H] 
    \centering
    \includegraphics[width=0.8\linewidth,height=0.8\linewidth]{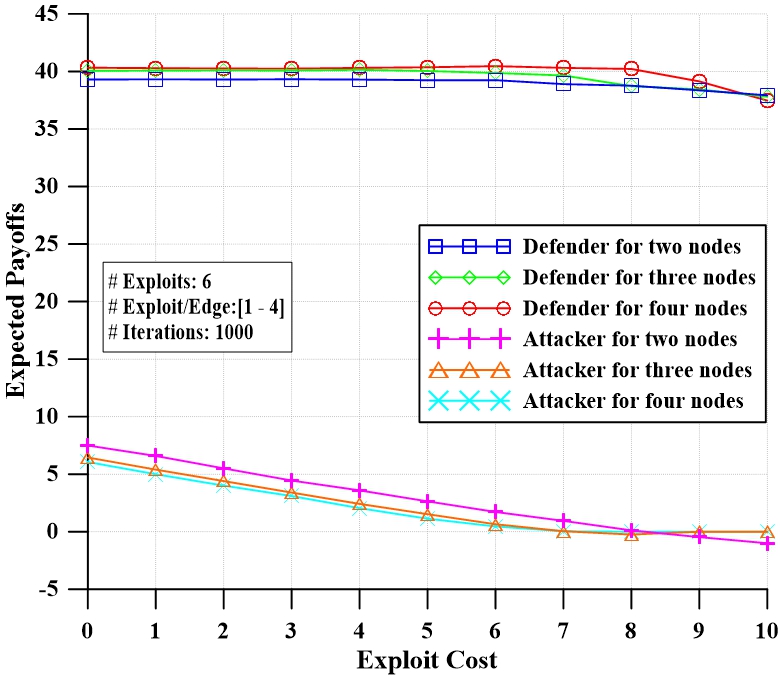} 
    \caption{Expected Payoffs with Varying Exploit Costs} 
    \label{fig:exploit_cost} 
\end{figure}

The impact of varying exploit costs on the expected payoffs for both the defender and the attacker is depicted in Figure~\ref{fig:exploit_cost}. This analysis helps us understand how changes in the cost of exploits influence strategic decisions and outcomes for both players in the security game.
Figure~\ref{fig:exploit_cost} shows the expected payoffs for both the defender and the attacker across different exploit costs, ranging from 0 to 10. The graph includes results for scenarios with two, three, and four nodes.

As the exploit cost increases, the defender's expected payoff shows a slight decreasing trend, but it remains relatively stable. This stability suggests that higher exploit costs slightly reduce the defender's advantage but do not drastically impact the overall payoff. For configurations with more nodes (three and four nodes), the expected payoffs for the defender remain consistently high, indicating the robustness of the defense strategy across different network sizes.

The expected payoffs for the attacker decrease significantly as the exploit cost increases. This trend indicates that higher exploit costs discourage attackers, reducing their expected payoff. The decrease in the attacker's payoff is more pronounced in smaller network configurations (two nodes). As the number of nodes increases, the attacker's expected payoffs are slightly higher but still follow a decreasing trend with increasing exploit costs.
The attacker's strategy is heavily influenced by exploit costs, with higher costs leading to less successful attacks and lower payoffs.
The difference in expected payoffs between different node configurations is more noticeable for the attacker. The attacker's expected payoff is lowest in the two-node scenario and highest in the four-node scenario.

For the defender, the expected payoffs are relatively similar across different node configurations, suggesting that the defense strategy is effective regardless of the number of nodes in the network.
\subsection{Scalability}

\begin{figure}[H] 
    \centering
    \includegraphics[width=0.8\linewidth,height=0.8\linewidth]{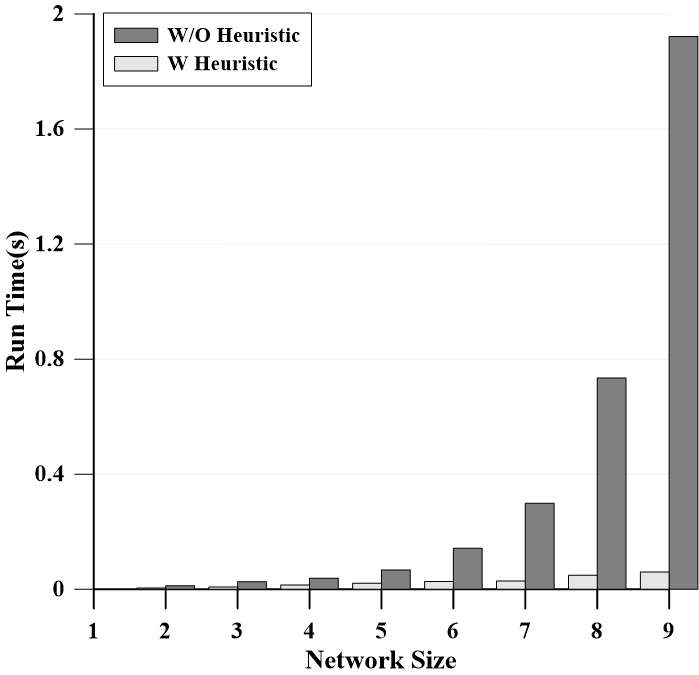} 
    \caption{Run-time comparison between with Heuristic and without Heuristic} 
    \label{fig:scalability} 
\end{figure}

Figure~\ref{fig:scalability} provides the run-time comparison between heuristic and non-heuristic approaches. The heuristic method significantly reduces run-time as the network size increases compared to the non-heuristic method. This efficiency is crucial for large-scale networks where computational resources and time are limited. The heuristic approach enables more rapid assessment and adjustment of strategies, ensuring timely and effective responses to potential threats.

The run-time comparison highlights the practical advantages of using heuristic methods in security games. Without heuristics, the time required to compute optimal strategies grows exponentially with the network size, making it impractical for larger networks. Conversely, the heuristic method maintains a relatively linear and low run-time, demonstrating its suitability for real-world applications where quick decision-making is essential.

\section{Conclusion}\label{sec:Conclusion}

In this study, we have developed a non-zero-sum game-theoretic model to analyze and derive optimal strategies for cyber defense. Our model takes into account the probabilities and costs associated with various exploits, the values of network nodes, and the costs of deploying defensive mechanisms such as honeypots. Through extensive simulations, we observed that the value of a node significantly affects the probability of different exploits being used. Higher value nodes are more likely to be targeted with high-success, low-cost exploits like Social Engineering and Phishing, while lower value nodes see an increased usage of Phishing and Malware.

Our findings demonstrate that defenders can benefit from understanding these attack patterns to better allocate their defensive resources. By prioritizing defenses on high-value nodes and anticipating the most likely exploits, defenders can enhance the overall security of the network. Additionally, our comparison between heuristic and non-heuristic approaches shows that heuristics do not significantly alter the optimal strategies but can improve computational efficiency.

Future research may extend this model to dynamic network environments, considering scenarios where multiple attackers utilize varied exploits, and node values and exploit costs fluctuate over time. Furthermore, incorporating more sophisticated attack and defense strategies and exploring the role of information asymmetry between attackers and defenders could provide deeper insights into cyber defense optimization. This study contributes significantly to the ongoing efforts to develop robust and effective cybersecurity strategies using game theory, offering practical guidance for resource allocation and strategic planning in cybersecurity.

\bibliographystyle{IEEEtran}
\bibliography{bibliography}
\vspace{12pt}
\color{red}

\end{document}